\documentclass[runningheads]{llncs}
\usepackage[T1]{fontenc}
\usepackage{graphicx}
\usepackage{graphicx,verbatim}
\usepackage{amsmath}
\usepackage{booktabs}
\usepackage{xcolor}
\usepackage{xcolor}
\usepackage{amsmath}
\usepackage{placeins}   

\usepackage{mathtools}
\usepackage{multirow}   

\begin{document}

\title{Brain Tumor Concentration Estimation: A Lightweight Optimization}
\title{A Lightweight Simulation and Learning Free Brain Tumor Concentration Estimation}

\title{How Butterfly Tumors Grow}

%
\title{How Butterfly Tumors Grow: DTI Based Tumor Simulation}

\title{From Fiber Tracts to Tumor Spread: Biophysical Modeling of Butterfly Glioma Growth Using Diffusion Tensor Imaging}

\author{Jonas Weidner\inst{1,2}  \and Ivan Ezhov\inst{1} \and Michal Balcerak\inst{3}\and André Datchev\inst{1} \and Lucas Zimmer\inst{1,2} \and Daniel Rueckert\inst{1,2,4} \and Björn Menze\inst{3} \and  Benedikt Wiestler\inst{1,2}}
\authorrunning{Weidner et al.}
\titlerunning{From Fiber Tracts to Tumor Spread}

\institute{Technical University of Munich\\ \and Munich Center for Machine Learning \\ \and University of Zurich\\ \and Imperial College London \\ \email{j.weidner@tum.de}}

\maketitle              

\begin{abstract}

Butterfly tumors are a distinct class of gliomas that span the corpus callosum, producing a characteristic butterfly-shaped appearance on MRI. The distinctive growth pattern of these tumors highlights how white matter fibers and structural connectivity influence brain tumor cell migration. To investigate this relation, we applied biophysical tumor growth models to a large patient cohort, systematically comparing models that incorporate fiber tract information with those that do not. Our results demonstrate that including fiber orientation data significantly improves model accuracy, particularly for a subset of butterfly tumors. These findings highlight the critical role of white matter architecture in tumor spread and suggest that integrating fiber tract information can enhance the precision of radiotherapy target volume delineation.

\end{abstract}

\begin{figure}[h!]
    \centering
    \includegraphics[width= \columnwidth]{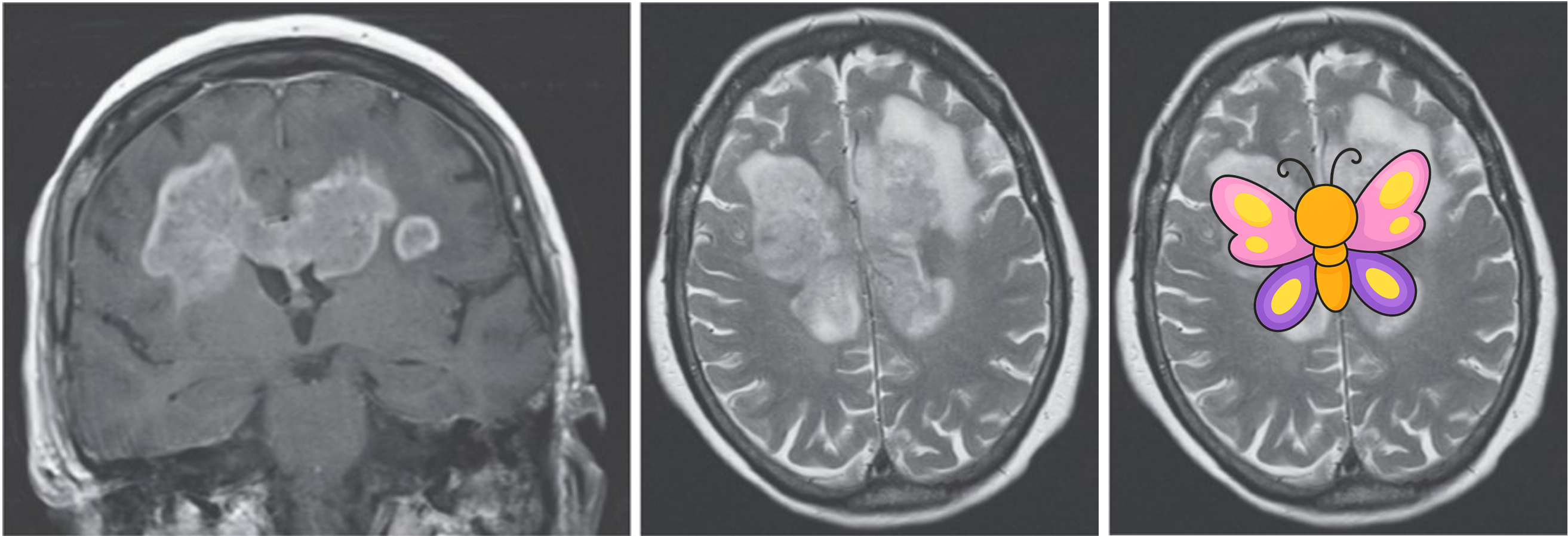}
    \caption{Butterfly tumors are gliomas spreading between both brain hemispheres. The infiltration follows the brain fiber tracts across the corpus callosum. (Figure adapted from \cite{Holzgreve2018Schmetterling})}
    \label{fig:graphicalAbstract}
\end{figure}

\section{Introduction}
Calling a lethal tumor a ‘butterfly’ is paradoxical. It places the gentle image of a beautiful, lively creature beside the harsh medical reality of a lethal disease.

In neuro-oncology, the “butterfly” label is descriptive: on MRI, an infiltrating glioblastoma can spread from one cerebral hemisphere to the other through the corpus callosum, producing two roughly symmetrical wings of tumor tissue bridging the brain’s midline. The tumor invades both hemispheres and uses the commissural fibers of the corpus callosum as a natural corridor for spread \cite{Bjorland2022Butterfly}. Population-based analysis consistently reports median overall survival below one year, which is significantly worse than typical glioblastoma \cite{Hazaymeh2022Prognostic}.

Experimental and clinical evidence shows that glioma cells migrate anisotropically, following the orientation of axonal bundles and other white matter structures \cite{bette2017local,rao2013mimicking,halperin1989radiation,zetterling2016extension}. Diffusion tensor imaging (DTI) maps the principal direction of water diffusion, which parallels axonal orientation in healthy tissue, and therefore provides an \textit{in vivo} proxy for the preferential paths of tumor spread. Quantitative analyses of fractional anisotropy and mean diffusivity demonstrate that infiltration fronts align with major tracts such as the corpus callosum, corona radiate, and cingulum \cite{Latini2021WhiteMatter}. 

Mathematical descriptions of glioma growth have long incorporated diffusion terms to simulate tumor infiltration. The typical approach to include tumor heterogeneity in brain tissue is to overweigh the diffusion within white matter compared to gray matter \cite{Swanson2000DifferentialMotility}. However, these models do not include anisotropic diffusion, but rather simply assume isotropic diffusion. 

Successive work embedded the full diffusion tensor into reaction–diffusion equations, creating anisotropic models \cite{Swan2018Anisotropic}. A slight improvement in patient fit was found by including fiber information. However, only the enhancing part of the tumor was considered, while the surrounding edema was ignored. The growth and diffusion rates were fixed for all patients based on empirical findings and not inferred individually, and only one tumor in one hemisphere was considered, no butterfly tumors. In total, only 10 patients were investigated.

Recently, a hybrid framework that combines shortest path algorithms with classical diffusion to approximate microscopic tumor fronts was presented \cite{bortfeld2022modeling}. By embedding patient-specific diffusion tensors, their model produced elongated, asymmetric invasion patterns that followed white matter orientation more closely than isotropic diffusion alone, and the anisotropic margins derived from these paths reduced normal tissue irradiation while preserving coverage of high-risk regions. However, the prediction is purely path-based and does not consider growth dynamics. 

Here, we investigate the "added value" of integrating fibre information into state-of-the-art tumor growth models to study and exploit the relation between fibre orientation and tumor spread in the brain. Our contributions are the following:
\begin{itemize}

\item We propose a novel way to include fiber information into personalized reaction-diffusion tumor growth models.
\item We systematically study patient-specific anisotropic reaction-diffusion models in a large patient cohort.
\item We demonstrate that adding fiber orientation improves the model fit, strongest for tumors crossing the corpus callosum, the butterfly tumors.




\end{itemize}

\section{Methods}
We utilize a biophysical tumor model to describe the growth process of brain tumors. Based on a simple imaging model, we fit the model to individual patients using a state-of-the-art evolutionary algorithm. 
\subsection{Biophysical Tumor Model}
We use the well-established Fisher-Kolmogorov equation to model brain-tumor growth. This model captures the spatiotemporal dynamics of glioma progression through proliferation and diffusion. It enables subject-specific simulation by incorporating anatomical priors and diffusion anisotropy.

\label{sec:brain_tumour_sim}

\begin{equation}
  \frac{\partial c(\mathbf{x},t)}{\partial t}
  \;=\;
  \nabla \!\cdot\!\bigl(D(\mathbf{x})\,\nabla c(\mathbf{x},t)\bigr)
  \;+\;
  \rho\,c(\mathbf{x},t)\,\bigl(1 - c(\mathbf{x},t)\bigr),
  \label{eq:fisher_kolmogorov}
\end{equation}
where $c(\mathbf{x},t)$ is the tumour‐cell density depending on time $t$ and location $x$. 
$D(\mathbf{x})$ it the tissue‐dependent diffusion tensor, and
$\rho$ is the proliferation rate.

We can separate the diffusion tensor into an isotropic part and an anisotropic part.  

\begin{equation}
D = D_\text{isotropic} + D_\text{anisotropic}
\end{equation}

The isotropic diffusion coefficient $D_\text{isotropic}$ is characterized by the scalar diffusion coefficient $\delta$ and the ratio $r$ of white matter (WM), gray matter (GM) and cerebrospinal fluid (CSF) at each location. Thus, $\forall  x: r_\text{WM}(x) + r_\text{GM}(x) + r_\text{CSF}(x) = 1 $. The diffusion is assumed to be smaller in GM because tumour cells infiltrate GM more slowly and are assumed not to invade CSF \cite{ezhov2023learn}.

\begin{equation}
D_\text{isotropic}(x) =  \delta \\ ( r_\text{WM}(x) + 0.1 \cdot r_\text{GM}(x)) \\ I
\end{equation}

$I$ represents the identity matrix. 

For the anisotropic part, we propose a novel way to include fiber information extracted from DTI. We assume a linear relation between measured DTI and tumor diffusion. Therefore, we normalize the raw DTI tensor inside the white matter brain mask to get the standard normally distributed tensor $D_\text{DTI}$. This is scaled by the anisotropic coefficient $\gamma$. The gray matter diffusion is assumed to be isotropic. 

\begin{equation}
D_\text{anisotropic}(x) = \gamma D_\text{DTI}(x)
\end{equation}

Note that this formulation allows the fallback into the isotropic case for vanishing $\gamma$: $\lim_{\gamma \to 0}  D \to D_\text{isotropic}$.

\begin{figure}[h!]
    \centering
    \includegraphics[width= \columnwidth]{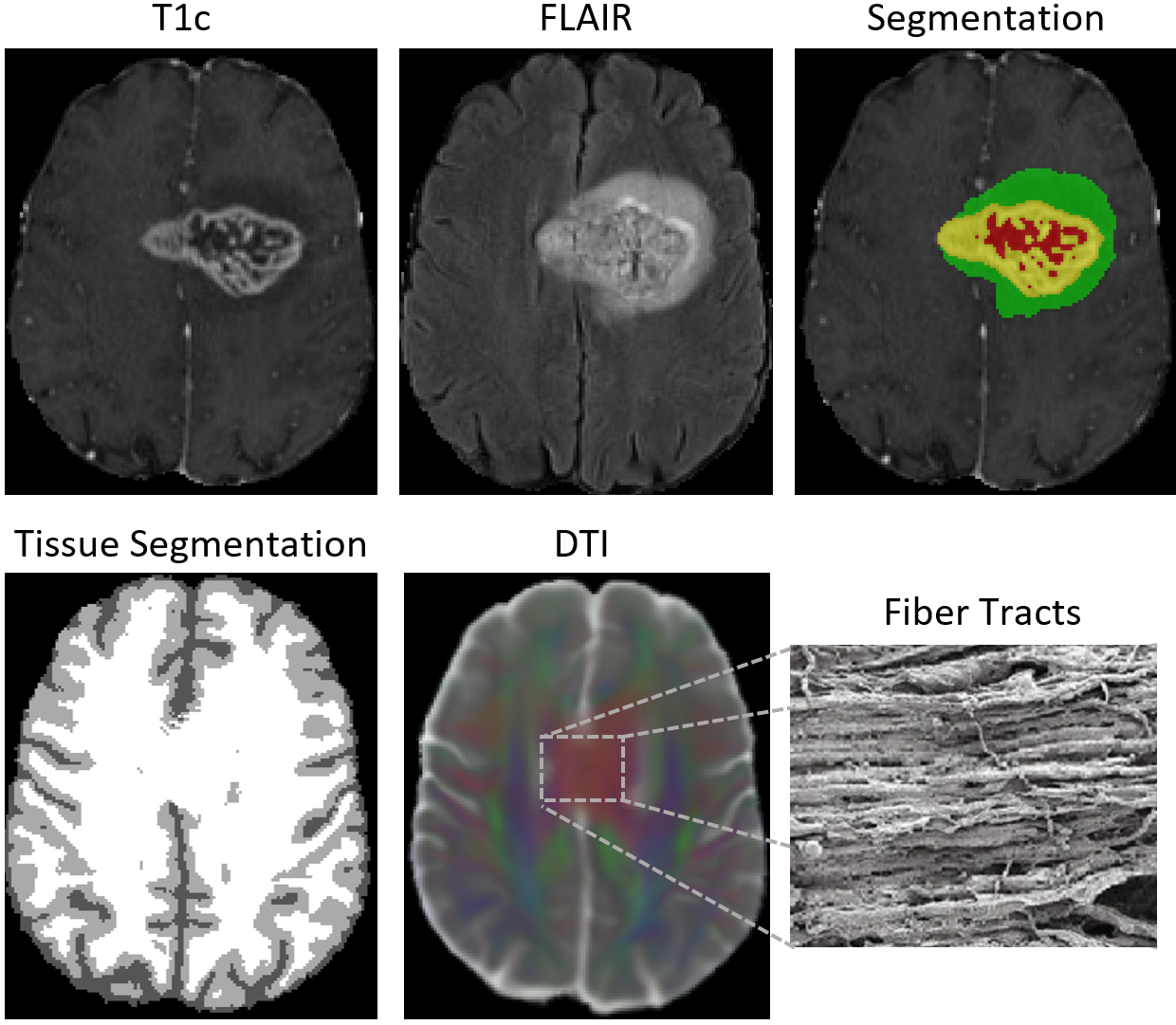}
    \caption{Preprocessing example of DTI registration. The measured MRIs are shown in the first row. In the second row, we register an atlas brain anatomy onto the patient to get a healthy version of the brain, with the anatomy of the final state. Thus, we can extract pre-tumor tissue segmentation and diffusion tensor imaging (DTI) fiber information. The color coding in the DTI describes the fiber orientation. The corpus callosum (red), connecting both hemispheres of the brain, has the strongest anisotropy due to the dense fibers.}
    \label{fig:preprocess}
\end{figure}
\subsection{Data}
We used the publicly available BraTS dataset \cite{menze2014multimodal}. We randomly selected 100 single-hemisphere tumors for the baseline cohort. We also curated eight butterfly-tumor cases for focused analysis.

For the biophysical model, we need the original, pre-tumor brain anatomy to reconstruct the growth process (Figure \ref{fig:preprocess}). As these healthy images are typically not available, and the grown tumor in the preoperative image does not allow for a reasonable segmentation of tissue, we register the Human Connectome Project (HPC) atlas \cite{Yeh2022_HCP1065} T1, which was preregistered to the Montreal Neurological Institute (MNI) atlas\cite{fonov2011unbiased}, into the patient space. This provides a first-order approximation of the healthy brain, but in the anatomical shape of the final state, which allows us to ignore the mass effect and simulate in a steady tissue state. For this, we use the ANTs algorithm \cite{tustison2010n4itk} to register the T1c image (which typically has the highest resolution) to the atlas. With the same deformation field, we also register the tissue segmentation and the DTI. It is important to mention that for the registration of the DTI, we also need to rotate the tensors according to the deformation field. Therefore, we used the \textit{ReorientTensorImage} function from the ANTs library \cite{tustison_antsx_2021}.

This registration provides an estimate of the fiber tracts also for patients without a DTI measurement. We also use this registered image for patients where there would be a DTI available, as the tumor destroys the underlying fibers.

Additionally, we can partly adjust for the mass effect, which describes the tumor-induced tissue shift by using the deformation field from the registration as an estimate.

 \begin{figure}[h!]
    \centering
    \includegraphics[width= 0.7 \columnwidth]{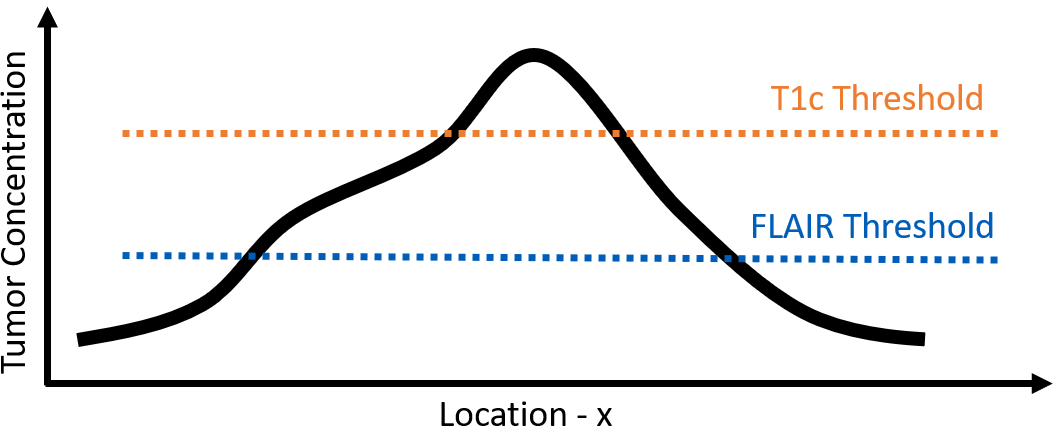}
    \caption{The imaging functions describe the mapping between continuous tumor concentration and MRI. In this study, we use a commonly used imaging function that is based on the T1c and FLAIR tumor segmentation. We assume that the threshold for the tumor concentration to be visible in the FLAIR image is $0.33$ and $0.66$, respectively, in the T1c image.}
    \label{fig:img_model}
\end{figure}

\subsection{Evaluation Metrics}
We assess segmentation accuracy with the Dice score. Based on our imaging function (Figure \ref{fig:img_model}), we binarize the continuous tumor concentration to compare it with the measured MRI. We calculate the Dice for the edema and tumor core, which is the necrosis and enhancing part, separately. Additionally, we compare a volume-weighted combined Dice. 

\subsection{Fitting the Model to Patients}
For individualized therapy, the biophysical model is fitted to a specific patient. The forward solver requires the proliferation rate $\rho$, diffusion
coefficient $\delta$, DTI coefficient $\gamma$, seed location $(x,y,z)$, and total growth time $T$,
i.e.\ the parameter set

\begin{equation}
    \theta_\mathrm{orig} = \{x,\,y,\,z,\,\rho,\,\delta,\gamma,\,T\}.
\end{equation}

With only a single clinical scan, $\theta$ is not identifiable. Thus, we normalize the final time and optimize $\theta_\mathrm{DTI} \coloneqq \theta_\mathrm{orig} \backslash \{T\} = \{x,\,y,\,z,\,\rho,\,\delta,\gamma \}$ instead. For the comparison to the isotropic case, we set  $\gamma = 0$ and optimize $\theta_\mathrm{isotropic} \coloneqq \theta_\mathrm{DTI} \backslash \{\gamma \} = \{x,\,y,\,z,\,\rho,\,\delta\}$. For the optimization process, we utilize covariance matrix adaptation - evolutionary strategy (CMA-ES) \cite{Weidner2024} and \textsc{TumorGrowthToolkit} \cite{balcerak2023individualizing} as a forward solver. As a loss function, we use the volume-weighted Dice between edema and T1c segmentation. 

\section{Results}

\begin{figure}[h!]
    \centering
    \includegraphics[width= 0.9\columnwidth]{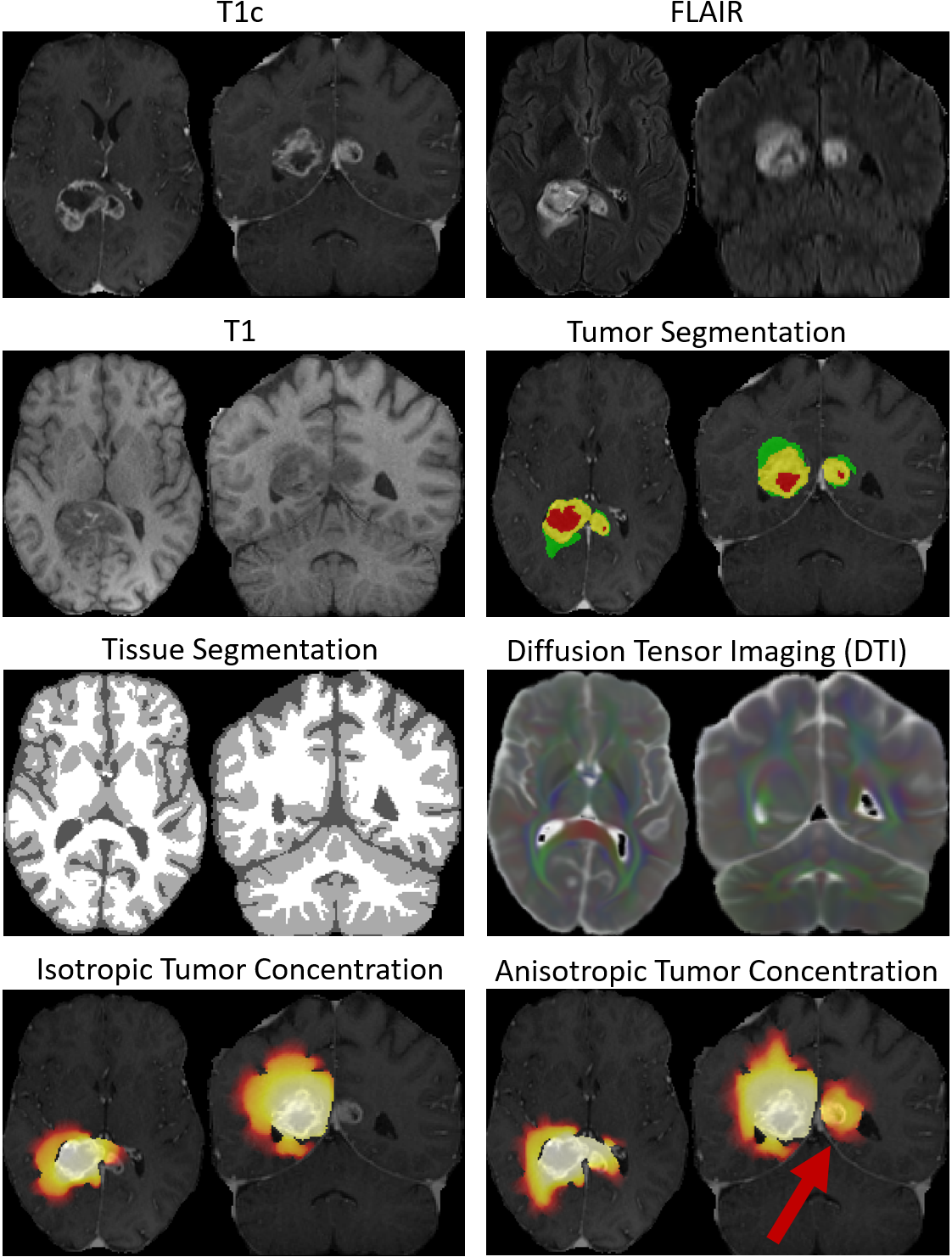}
    \caption{ Example of a butterfly tumor. We show the measured MRI modalities T1c, T1, FLAIR, and the tumor segmentation. In the next row, we display the registered tissue segmentation and DTI, which are used for the simulation. In the last row, we compare the isotropic model with the anisotropic model. We find that only the anisotropic model is able to reproduce the butterfly tumor growth behaviour properly (red arrow).}
    \label{fig:exampleRes}
\end{figure}

A qualitative example of tumor concentration estimation is shown in Figure \ref{fig:exampleRes}. We find that the anisotropic model outperforms the isotropic model. The anisotropic model approximates butterfly tumors much more accurately, in particular, capturing the transcallosal growth.

These qualitative findings are backed by the quantitative results shown in Table \ref{tab:res}.  In the large study on single hemispheric tumors, we find that the Dice score is improved by about $1.2\%$ for tumor core and edema in the anisotropic model over the isotropic model. This finding suggests that incorporating fiber information generally improves tumor-growth modeling.

The improvement of the anisotropic model is especially visible for the butterfly tumors. In the butterfly cohort, the Dice score for the tumor core rises by roughly $5\%$. This underlines the assumption of the strong dependence on fiber information for butterfly tumors.

\begin{table}[h]
\caption{Mean Dice scores with standard error. Comparing the anisotropic and isotropic biophysical models, as well as the pairwise improvement of the anisotropic model on a large cohort of single-hemisphere tumors and butterfly tumors. We find significant improvement ("$ * $" for $p<0.05$ and "$ ** $" for $p<0.01$ using a paired t-test), for the anisotropic approach.}
\label{tab:res}
\centering
\begin{tabular}{lcccc}
\hline
Tumor Type           & Model               & Weighted Dice & Edema & Tumor Core \\
\hline

\multirow{3}{*}{Single Hemisphere}
                     & Isotropic             & $0.613 \pm 0.010$ & $0.506 \pm 0.014$ & $0.721 \pm 0.012$ \\
                     & Anisotropic            & $0.626 \pm 0.009^{**}$ & $0.519 \pm 0.014^{**}$ & $0.733 \pm 0.010^{*}$ \\
                     & Pairwise Improvement  & $0.013 \pm 0.002^{**}$ & $0.013 \pm 0.003^{**}$ & $0.012 \pm 0.006^{*}$ \\
\hline

\multirow{3}{*}{Butterfly}
                     & Isotropic             & $0.527 \pm 0.042$ & $0.384 \pm 0.061$ & $0.669 \pm 0.028$ \\
                     & Anisotropic            & $0.550 \pm 0.039$ & $0.384 \pm 0.067$ & $0.716 \pm 0.021$ \\
                     & Pairwise Improvement  & $0.023 \pm 0.011$ & $0.000 \pm 0.011$ & $0.047 \pm 0.031$ \\

\hline


\end{tabular}
\end{table}

\section{Discussion}
Our results confirm that embedding atlas fiber orientation, warped to individual patients' anatomy, into the Fisher Kolmogorov model yields a significant gain in accuracy, with the effect being most pronounced in butterfly gliomas that traverse the corpus callosum. The modest but consistent improvement in single-hemisphere cases suggests that white matter anisotropy influences tumor growth in general. Thus, our results agree with the observations of \cite{Swan2018Anisotropic}, extending them to a statistically robust cohort. At the same time, the limited magnitude of the Dice increase indicates that other factors, such as microenvironmental heterogeneity or mass effect, still constrain predictive power. The registration-based recovery of healthy tissue and fibers might lead to minor errors, partially hindering the benefit of anisotropy. Future work should therefore focus on refining anatomy reconstruction and on jointly estimating patient-specific proliferation and diffusion parameters from longitudinal images to further tighten the model fit.

\section{Conclusion}
We introduced the first large-scale evaluation of an anisotropic reaction diffusion model for glioma that exploits diffusion tensor-derived fiber orientation. The approach yields statistically significant improvements over its isotropic counterpart, with the largest gains observed in butterfly tumors where commissural tracts dominate spread. These findings support the routine inclusion of fiber information in biophysical modeling and motivate further development toward patient-tailored therapy planning.

%
%
%
\FloatBarrier  
\bibliographystyle{splncs04}
\bibliography{bib.bib}
\end{document}